\documentstyle[twoside,fleqn,espcrc2,epsfig]{article}

\newcommand{\AmS}{{\protect\the\textfont2
  A\kern-.1667em\lower.5ex\hbox{M}\kern-.125emS}}

\title{Low mass lepton pair production in hadron collisions}

\author{E.L. Berger and M. Klasen \address{HEP Theory Group, Argonne National
        Laboratory, 9700 South Cass Avenue, Argonne, IL 60439, USA}\thanks{
        Supported by the U.S.\ Department of Energy, Division of High Energy
        Physics, under Contract W-31-109-ENG-38.}}

\begin{document}

\begin{abstract}
The hadroproduction of lepton pairs with mass $Q$ and transverse
momentum $Q_T$ can be described in perturbative QCD by the same partonic
subprocesses as prompt photon production. We demonstrate that, like prompt
photon production, lepton pair production is dominated by quark-gluon
scattering
in the region $Q_T>Q/2$. This leads to sensitivity to the gluon density
in kinematical regimes that are accessible both at collider and fixed target
experiments while eliminating the theoretical and experimental uncertainties
present in prompt photon production.
\end{abstract}

\maketitle

\vspace*{-7.5cm} \noindent ANL-HEP-CP-99-72
\vspace*{ 6.2cm}

\section{INTRODUCTION}
The production of lepton pairs in hadron collisions $h_1h_2\rightarrow\gamma^*
X;\gamma^*\rightarrow l\bar{l}$ proceeds through an intermediate
virtual photon and its subsequent leptonic decay. Traditionally, interest in
this Drell-Yan process has concentrated on lepton pairs with
large mass $Q$ which allows for the application of perturbative QCD and the
extraction of the antiquark density in the proton \cite{Drell:1970wh}.

Prompt photon production $h_1h_2\rightarrow\gamma X$ can be calculated in
perturbative QCD if the transverse momentum $Q_T$ of the photon is
sufficiently large. This process then provides essential information on the
gluon density in the proton at large $x$ \cite{Martin:1998sq}. Unfortunately,
it suffers from considerable fragmentation, isolation, and intrinsic transverse
momentum uncertainties. Alternatively, the gluon density can be constrained
from the production of jets with large transverse momentum at hadron colliders
\cite{Lai:1999wy}, which however suffers from ambiguous information coming
from different experiments and colliders.

In this paper we demonstrate that, like prompt photon production, lepton pair
production is dominated by quark-gluon scattering in the region $Q_T>Q/2$.
This leads to sensitivity to the gluon density in kinematical regimes
that are accessible both at collider and fixed target experiments while
eliminating the theoretical and experimental uncertainties.

In Sec.~\ref{sec:2}, we briefly discuss the relationship between virtual and
real photon production in hadron collisions in next-to-leading order QCD.
In Sec.~\ref{sec:3} we present our numerical results, and Sec.~\ref{sec:4}
contains a summary.

\section{NEXT-TO-LEADING ORDER QCD FORMALISM}
\label{sec:2}

In leading order (LO) QCD, two partonic subprocesses contribute to the
production of virtual and real photons with non-zero transverse momentum:
$q\bar{q}\rightarrow\gamma^{(*)}g$ and $qg\rightarrow\gamma^{(*)}q$.
The cross section for lepton pair production is related to the cross section
for virtual photon production through the leptonic branching ratio of the
virtual photon $\alpha/(3\pi Q^2)$. The virtual photon cross section reduces
to the real photon cross section in the limit $Q^2\rightarrow 0$.

The next-to-leading order (NLO) QCD corrections arise from virtual one-loop
diagrams interfering with the LO diagrams and from real emission diagrams. At
this order processes with incident gluon pairs $(gg)$, quark pairs $(qq)$,
and non-factorizable quark-antiquark $(q\bar{q}_2)$ processes contribute also.
Singular contributions are regulated in $n=4-2\epsilon$ dimensions and
removed through $\overline{\rm MS}$ renormalization, factorization, or
cancellation between virtual and real contributions. An important difference
between virtual and real photon production arises when a quark emits a
collinear photon. Whereas the collinear emission of a real photon leads to a
$1/\epsilon$ singularity that has to be factorized into a fragmentation
function, the collinear emission of a virtual photon gives a finite
logarithmic contribution since it is regulated naturally by the photon
virtuality $Q$. In the limit $Q^2\rightarrow 0$ the NLO virtual photon
cross section reduces to the real photon cross section if this logarithm is
replaced by a $1/\epsilon$ pole. A more detailed discussion can be found
in \cite{Berger:1998ev}.

The situation is completely analogous to hard
photoproduction where the photon participates in the scattering in the initial
state instead of the final state. For real photons, one encounters an
initial-state singularity that is factorized into a photon structure function.
For virtual photons, this singularity is replaced by a logarithmic dependence
on the photon virtuality $Q$ \cite{Klasen:1998jm}.

\section{NUMERICAL RESULTS}
\label{sec:3}

In this section we present numerical results for the production of lepton pairs
in $p\bar{p}$ collisions at the Tevatron with center-of mass energy
$\sqrt{S}=1.8$ and 2.0 TeV and in $pH^2$ collisions at fixed target experiments
with $\sqrt{S}=38.8$ GeV. We analyze the invariant cross section $Ed^3\sigma/
dp^3$ averaged over the rapidity interval -1.0 $<y<$ 1.0 at the Tevatron and
averaged over the scaled longitudinal momentum interval 0.1 $< x_F <$ 0.3 at
fixed target experiments. We integrate the cross section over various
intervals of $Q$ and plot it as a function of the transverse momentum $Q_T$.
Our predictions are based on a NLO QCD calculation \cite{Arnold:1991yk} and
are evaluated in the $\overline{\rm MS}$ renormalization scheme. The
renormalization and factorization scales are set to $\mu=\mu_f=
\sqrt{Q^2+Q_T^2}$. If not stated otherwise, we use the CTEQ4M
parton distributions \cite{Lai:1997mg} and the corresponding value of
$\Lambda$ in the two-loop expression of $\alpha_s$ with four flavors (five if
$\mu>m_b$). The Drell-Yan factor $\alpha/(3\pi Q^2)$ for the decay of the
virtual photon into a lepton pair is included in all numerical results.

\begin{figure}[htb]
 \begin{center}
  {\unitlength1cm
  \begin{picture}(7.6,10.5)
   \epsfig{file=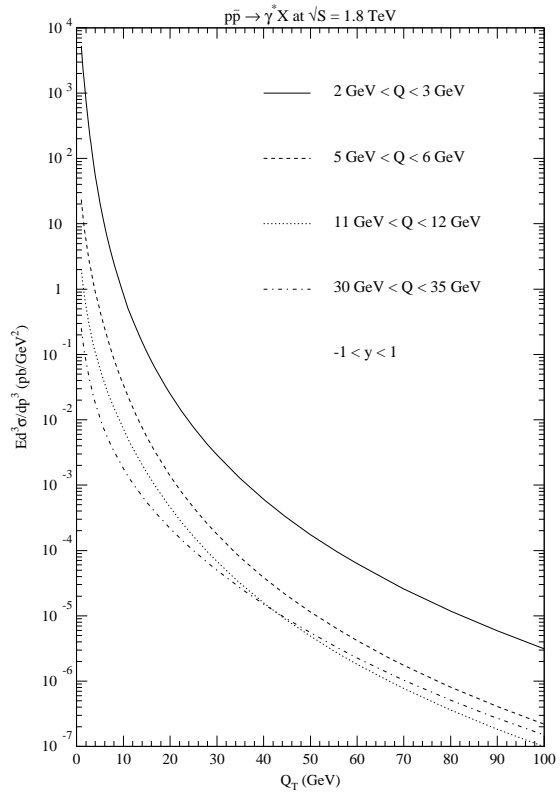,bbllx=60pt,bblly=100pt,bburx=495pt,bbury=725pt,%
           height=10.5cm}
  \end{picture}}
 \end{center}
\vspace*{-1cm}
\caption{Invariant cross section $Ed^3\sigma/dp^3$ as a function of $Q_T$
for $p\bar{p} \rightarrow \gamma^* X$ at $\sqrt{S}=1.8$ TeV in
non-resonance regions of $Q$. The cross section falls with the mass of the
lepton pair $Q$ and, more steeply, with its transverse momentum $Q_T$.}
\label{fig:1}
\end{figure}

In Fig.~\ref{fig:1} we display the NLO QCD cross section for lepton pair
production at the Tevatron at $\sqrt{S}=1.8$ TeV as a function of $Q_T$ for
four regions of $Q$. The regions of $Q$ have been chosen carefully to avoid
resonances, {\it i.e.\ } between the $\rho$ and the $J/\psi$ resonances,
between the $J/\psi$ and the $\Upsilon$ resonances, above the $\Upsilon$'s,
and a high mass region. The cross section falls both with the mass of the
lepton pair $Q$ and, more steeply, with its transverse momentum $Q_T$.
Unfortunately, no data are available yet from the CDF and D0 experiments.
However, data exist for
prompt photon production out to $Q_T\simeq 100$ GeV, where the cross section
is about $10^{-3}$ pb/GeV$^2$. It should therefore be possible to analyze Run I
data for lepton pair production up to at least $Q_T\simeq 30$ GeV where one
can probe the parton densities in the proton up to $x_T = 2Q_T/\sqrt{S}\simeq
0.03$. The UA1
collaboration measured the transverse momentum distribution of lepton
pairs at $\sqrt{S}=630$ GeV up to $x_T=0.13$ \cite{Albajar:1988iq}, and their
data agree well with our theoretical results \cite{Berger:1998ev}.

\begin{figure}[htb]
 \begin{center}
  {\unitlength1cm
  \begin{picture}(7.6,10.5)
   \epsfig{file=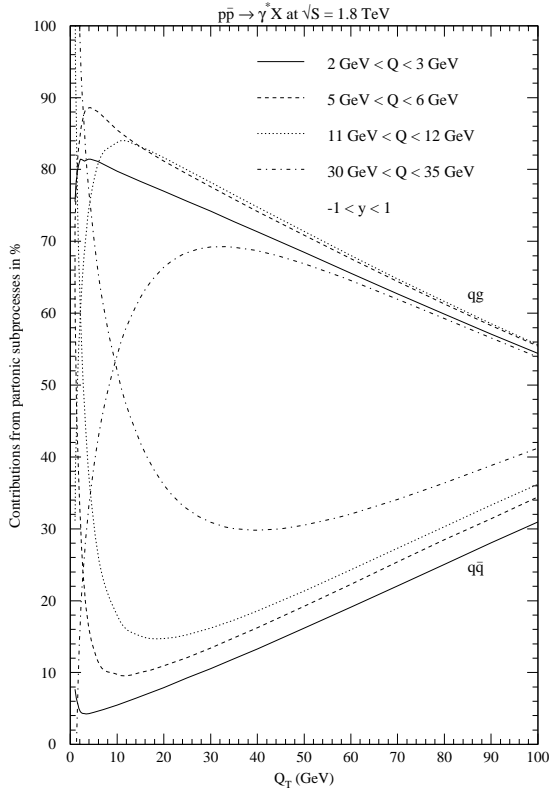,bbllx=60pt,bblly=100pt,bburx=495pt,bbury=725pt,%
           height=10.5cm}
  \end{picture}}
 \end{center}
\vspace*{-1cm}
\caption{Contributions from the partonic subprocesses $qg$ and $q\bar{q}$ to
the invariant cross section $Ed^3\sigma/dp^3$ as a function of $Q_T$
for $p\bar{p}\rightarrow \gamma^* X$ at $\sqrt{S}$ = 1.8 TeV. The
$qg$ channel clearly dominates in the region $Q_T > Q/2$.}
\label{fig:2}
\end{figure}

The fractional contributions from the $qg$ and $q\bar{q}$ subprocesses up
through NLO are shown in Fig.~\ref{fig:2}. It is evident from Fig.~\ref{fig:2}
that the $qg$ subprocess is the most important subprocess as long as
$Q_T > Q/2$. The dominance of the $qg$ subprocess diminishes somewhat with $Q$,
dropping from over 80 \% for the lowest values of $Q$ to about 70 \%
at its maximum for $Q \simeq$ 30 GeV. In addition, for very large $Q_T$, the
significant luminosity associated with the valence dominated $\bar{q}$
density in $p\bar{p}$ reactions begins to raise the fraction of the cross
section attributed to the $q\bar{q}$ subprocesses.

\begin{figure}[htb]
 \begin{center}
  {\unitlength1cm
  \begin{picture}(7.6,10.5)
   \epsfig{file=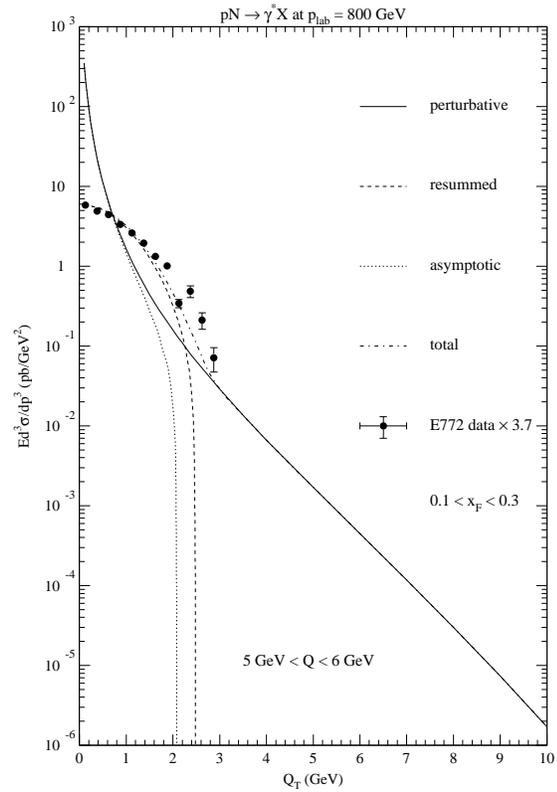,bbllx=60pt,bblly=100pt,bburx=495pt,bbury=725pt,%
           height=10.5cm}
  \end{picture}}
 \end{center}
\vspace*{-1cm}
\caption{Invariant cross section $Ed^3\sigma/d p^3$ as a function of 
$Q_T$ for $p N \rightarrow \gamma^* X$ at $p_{\rm lab}=$ 800 GeV in the
region between the $J/\psi$ and $\Upsilon$ resonances. The NLO
perturbative cross section (solid) is shown along with the all-orders resummed
expectation (dashed), the asymptotic result (dotted), and a matched expression
(dot-dashed). The data are from the Fermilab E772 collaboration.}
\label{fig:3}
\end{figure}

Data obtained by the Fermilab E772 collaboration \cite{McGaughey:1994dx}
from an 800 GeV proton beam incident on a deuterium target are shown in
Fig.~\ref{fig:3} along with theoretical calculations.
For our analysis we have
chosen a lepton pair mass region between the $J/\psi$ and $\Upsilon$
resonances.
The solid line shows the purely perturbative NLO expectation.
The transition to low $Q_T$ can be described by the
soft-gluon resummation formalism and is shown in the dashed curve in
Fig.~\ref{fig:3} \cite{Collins:1985kg,Ladinsky:1994zn}. The resummed result can
be expanded in a power series in $\alpha_s$ asymptotically around $Q_T=0$.
Its NLO component (dotted curve) can then be matched to the
perturbative result (dot-dashed curve) \cite{Arnold:1991yk}. From
Fig.~\ref{fig:3} it becomes clear that resummation is not needed and fixed
order
perturbation theory can be trusted when $Q_T>Q/2$. Unfortunately, the data
from E772 do not extend into this region. However, the cross section should
be measurable in forthcoming experiments
down to $10^{-3}$ pb/GeV$^2$, {\it i.e.} out to at least
$Q_T\simeq 6$ GeV or $x_T \simeq 0.31$ where the gluon density is poorly
constrained now.

\begin{figure}[htb]
 \begin{center}
  {\unitlength1cm
  \begin{picture}(7.6,10.5)
   \epsfig{file=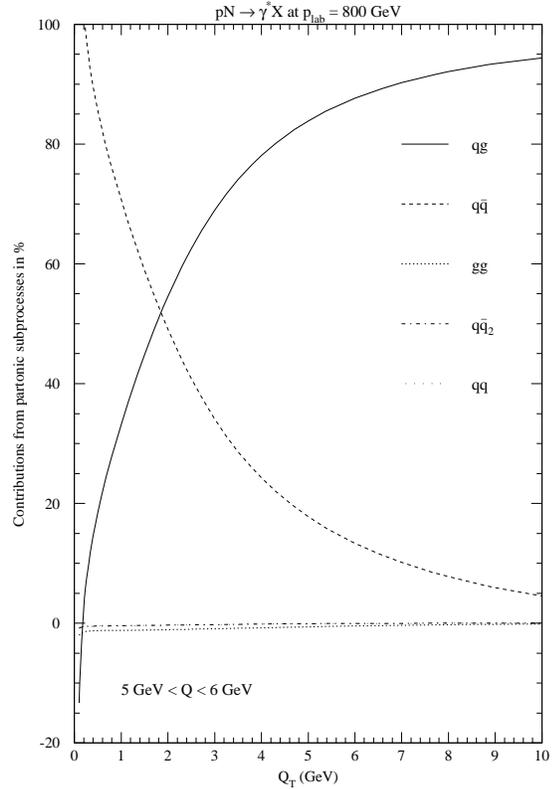,bbllx=60pt,bblly=100pt,bburx=495pt,bbury=725pt,%
           height=10.5cm}
  \end{picture}}
 \end{center}
\vspace*{-1cm}
\caption{Contributions from the NLO QCD partonic subprocesses to the invariant
cross section $Ed^3\sigma/dp^3$ as a function of $Q_T$ for $p N \rightarrow
\gamma^* X$ at $p_{\rm lab}$ = 800 GeV. $qg$ (solid) dominates over $q\bar{q}$
(dashed) at the level of 80 \% once $Q_T\simeq Q$, and the pure NLO QCD
processes $gg$ (dotted), $q\bar{q}_2$ non-factorizable parts (dot-dashed), and
$qq$ (wide dots) are negligible.}
\label{fig:4}
\end{figure}

In Fig.~\ref{fig:4} we demonstrate that in fixed target experiments also the
lepton pair cross section is dominated by quark-gluon scattering at the level
of 80 \% once $Q_T \simeq Q$. The results in Fig.~\ref{fig:4} also prove that
subprocesses other than those initiated by the $q\bar{q}$ and
$q g$ initial channels are of negligible import.

We will now turn to a previously unpublished study
of the sensitivity of collider and fixed target experiments to the gluon
density in the proton. The full uncertainty in the gluon density is not known.
Here we estimate this uncertainty from
the variation of different recent parametrizations. We choose the latest
global fit by the CTEQ collaboration (5M) as our point of reference
\cite{Lai:1999wy} and compare it to their preceding analysis (4M
\cite{Lai:1997mg}) and to a fit with a higher gluon density (5HJ) intended to
describe
the CDF (and D0) jet data at large transverse momentum. We also compare to
global fits by MRST \cite{Martin:1998sq}, who provide three different sets
with a central, higher, and lower gluon density, and to GRV98
\cite{Gluck:1998xa}\footnote{In this set a purely perturbative generation of
heavy flavors (charm and bottom) is assumed. Since we are working in a massless
approach, we resort to the GRV92 parametrization for the charm contribution
\cite{Gluck:1992ng} and assume the bottom contribution to be negligible.}.
For this study we update the Tevatron center-of-mass
energy to Run II conditions ($\sqrt{S}= 2.0$ TeV) which increases the invariant
cross section for the production of lepton pairs with mass 5 GeV $<Q<$ 6 GeV by
5 \% at low $Q_T \simeq 1$ GeV and 20 \% at high $Q_T \simeq 100$ GeV. 

\begin{figure}[htb]
 \begin{center}
  {\unitlength1cm
  \begin{picture}(7.6,10.5)
   \epsfig{file=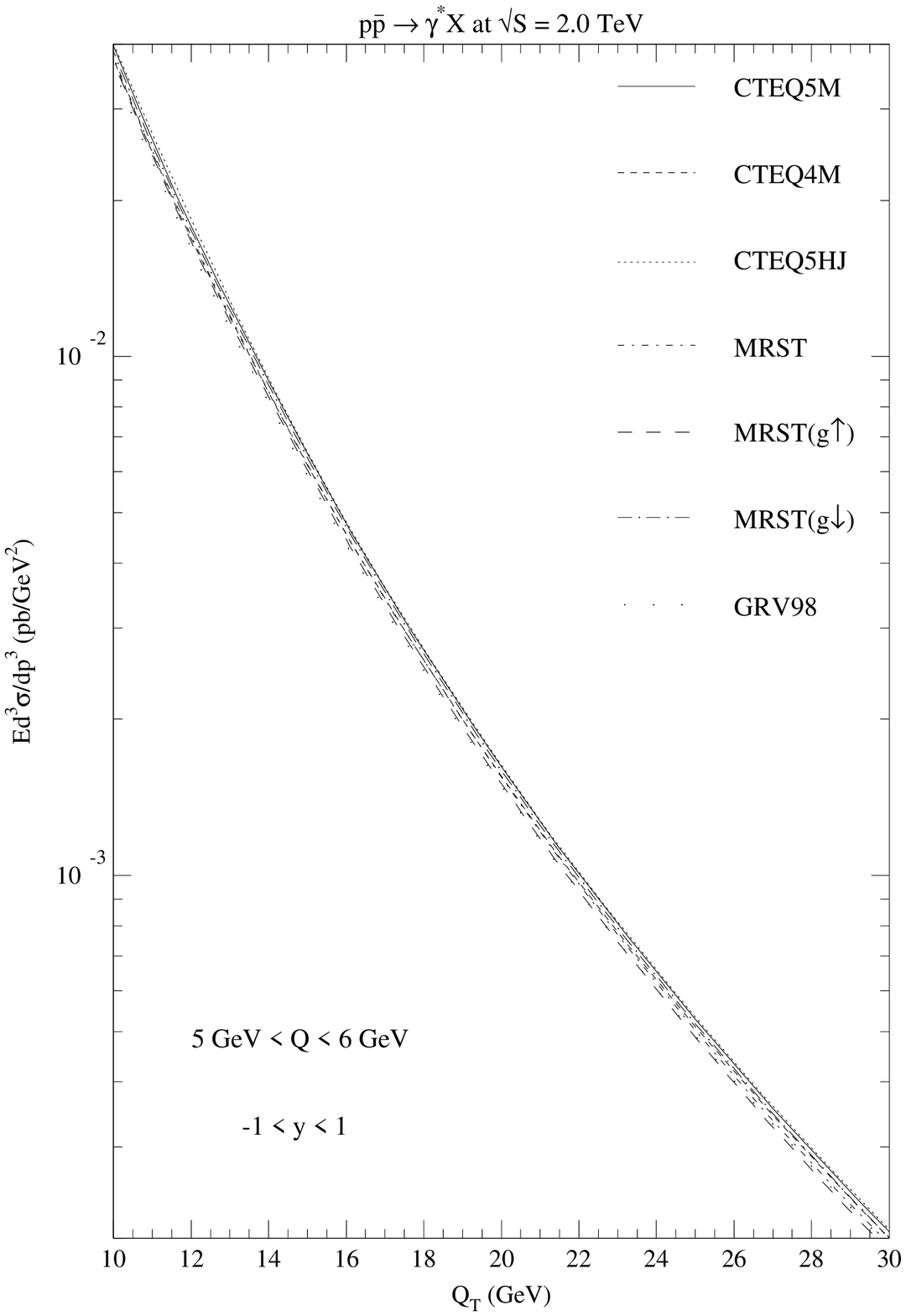,bbllx=60pt,bblly=100pt,bburx=495pt,bbury=725pt,%
           height=10.5cm}
  \end{picture}}
 \end{center}
\vspace*{-1cm}
\caption{Invariant cross section $Ed^3\sigma/dp^3$ as a function of $Q_T$
for $p\bar{p} \rightarrow \gamma^* X$ at $\sqrt{S}=2.0$ TeV in the
region between the $J/\psi$ and $\Upsilon$ resonances. The largest differences
from CTEQ5M are obtained with GRV98 at low $Q_T$ (minus 10 \%) and with
MRST(g$\uparrow$) at large $Q_T$ (minus 7 \%).}
\label{fig:5}
\end{figure}

In Fig.~\ref{fig:5} we plot the cross section for lepton pairs between the
$J/\psi$ and $\Upsilon$ resonances at Run II of the Tevatron which should be
measurable up to at least $Q_T\simeq 30$ GeV ($x_T \simeq 0.03$). For the CTEQ
parametrizations we find that the cross section increases from 4M to 5M by 2.5
\% ($Q_T=30$ GeV) to 5 \% ($Q_T=10$ GeV) and from 5M to 5HJ by 1 \% in the
whole $Q_T$-range. The largest differences to CTEQ5M are obtained with GRV98 at
low $Q_T$ (minus 10 \%) and with MRST(g$\uparrow$) at large $Q_T$ (minus 7\%).

\begin{figure}[htb]
 \begin{center}
  {\unitlength1cm
  \begin{picture}(7.6,10.5)
   \epsfig{file=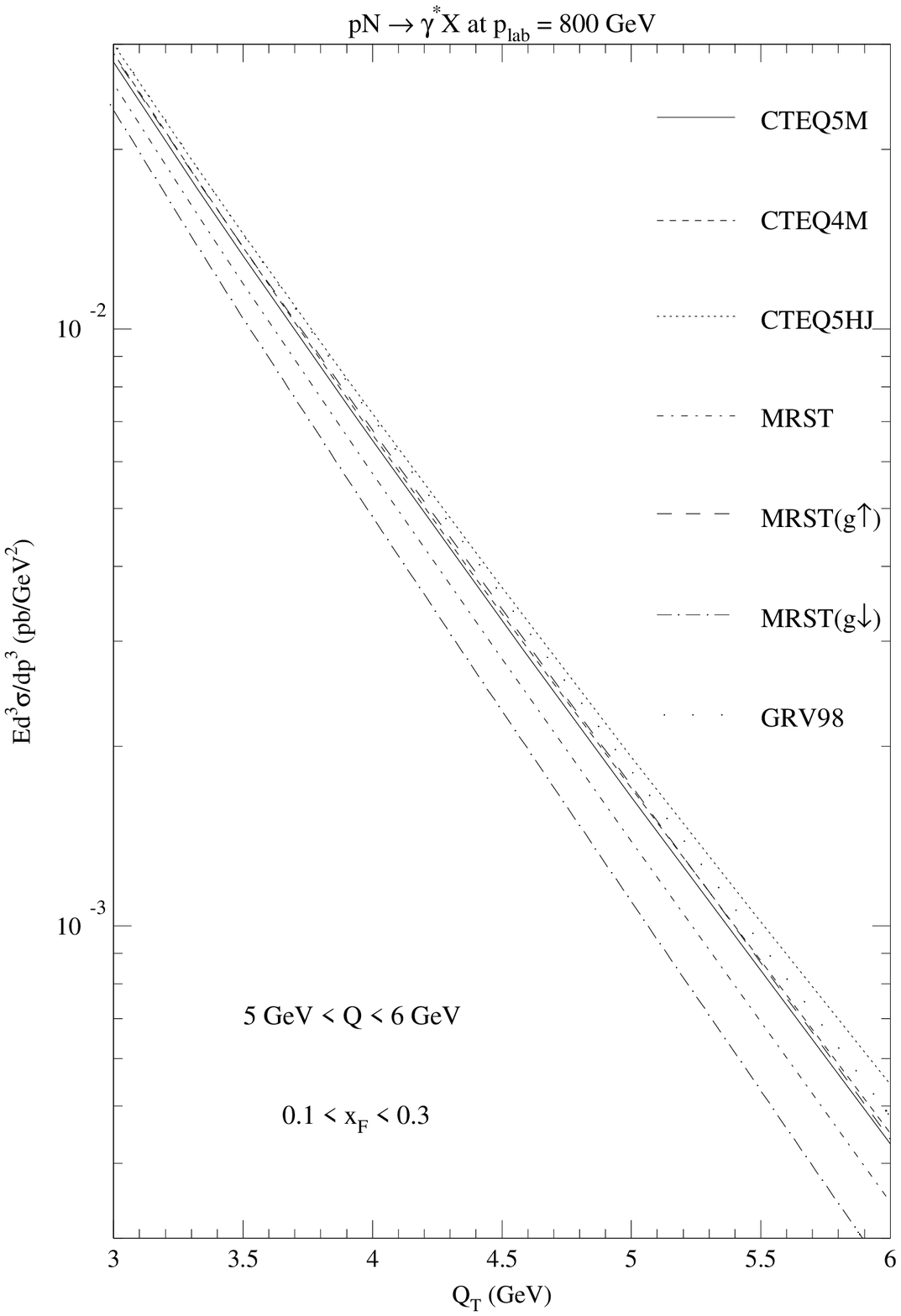,bbllx=60pt,bblly=100pt,bburx=495pt,bbury=725pt,%
           height=10.5cm}
  \end{picture}}
 \end{center}
\vspace*{-1cm}
\caption{Invariant cross section $Ed^3\sigma/d p^3$ as a function of 
$Q_T$ for $p N \rightarrow \gamma^* X$ at $p_{\rm lab}=$ 800 GeV. The cross
section is highly sensitive to the gluon distribution in the proton in regions
of $x_T$ where it is poorly constrained.}
\label{fig:6}
\end{figure}

A similar analysis for conditions as in Fermilab's E772 experiment is shown in
Fig.~\ref{fig:6}.
In fixed target experiments one probes substantially larger regions of $x_T$
than in collider experiments.
Therefore one expects a much larger sensitivity to the gluon distribution in
the proton. Indeed we find that CTEQ5HJ increases the cross section by 7 \%
(26 \%) w.r.t.\ CTEQ5M at $Q_T=3$ GeV ($Q_T=6$ GeV) and even by 134 \% at
$Q_T=10$ GeV. For MRST(g$\downarrow$) the CTEQ5M cross section drops by 17 \%,
40 \%, and 59 \% at these three values of $Q_T$. 

\section{SUMMARY}
\label{sec:4}
In summary, we have demonstrated that the production of Drell-Yan pairs with
low mass and large transverse momentum is dominated by gluon initiated
subprocesses. In contrast to prompt photon production, uncertainties coming
from fragmentation, isolation, and intrinsic transverse momentum are absent.
The hadroproduction of low mass lepton pairs is therefore an advantageous
source of information on the gluon density in the proton at large $x$ in
collider experiments and even more in fixed target experiments. Massive lepton
pair production data could provide new insights into the parametrization
and size of the gluon density.

\section*{Acknowledgement}
It is a pleasure to thank L.E.~Gordon for his collaboration.

\end{document}